\documentclass[titlepage,12pt,a4paper,reqno]{article}
\usepackage{amsmath,amssymb,amsfonts,graphicx,graphics}
\usepackage[linktocpage=true,colorlinks,pdftex]{hyperref}
\usepackage{xcolor}
\colorlet{linkequation}{blue}
\usepackage{bm}
\usepackage{doi}
\usepackage{hyperref}
\hypersetup{colorlinks, citecolor=violet, filecolor=black, linkcolor=black, urlcolor=blue}
\newcommand*{\refeq}[1]{%
  \begingroup
    \hypersetup{
      linkcolor=linkequation,
      linkbordercolor=linkequation,
    }%
    \ref{#1}%
  \endgroup
}
\allowdisplaybreaks

\begin{document} 


\begin{titlepage}

\centerline{\LARGE \bf Knots and the Maxwell Equations} 
\vskip 1cm
\centerline{ \bf Ion V. Vancea }
\vskip 0.5cm
\centerline{\sl Grupo de F{\'{\i}}sica Te\'{o}rica e Matem\'{a}tica F\'{\i}sica}
\centerline{\sl Departamento de F\'{\i}sica}
\centerline{\sl Universidade Federal Rural do Rio de Janeiro}
\centerline{\sl Serop\'{e}dica - Rio de Janeiro, Brazil}
\centerline{
\texttt{\small ionvancea@ufrrj.br} 
}

\vspace{0.5cm}

\centerline{25 March 2020}

\vskip 1.4cm
 
\centerline{\large\bf Abstract}
 
In this chapter, we review the Ra\~{n}ada field line solutions of Maxwell's equations in the vacuum, which describe a topologically non-trivial electromagnetic field, as well as their relation with the knot theory. Also, we present a generalization of these solutions to the non-linear electrodynamics recently published in the literature.

\vskip 0.7cm 

\noindent \textbf{Keywords:} { Maxwell's equations. Non-linear electrodynamics. Ra\~{n}ada solutions. Knot solutions.}
\noindent

\end{titlepage}


\section{Introduction}

The discovery of the knot solutions of Maxwell's equations in the vacuum represents one of the most exciting results obtained recently in the modern classical electrodynamics. Since their first appearance in the seminal papers \cite{Trautman:1977im,Ranada:1989wc,Ranada:1990,Ranada:1992hw}, many interesting properties, applications, and generalizations of the knot electromagnetic fields have been discovered. 

The knot solutions of Maxwell's equations can be described in terms of electric and magnetic field lines, and their topology can be given in terms of a pair of complex scalar fields that are interpreted as Hopf maps $S^3 \to S^2$ on the compactified space-like directions of the Minkowski space-time \cite{Ranada:1992hw}-\cite{Ranada:1997}. Among the properties studied up to now, one can cite: the relationship between the linked and the knotted electromagnetic fields discussed in \cite{Irvine:2008,Irvine:2010}, the dynamics of the electric charges in topologically non-trivial electromagnetic backgrounds investigated in \cite{Kleckner:2013}-\cite{Ranada:2017ore}, and the topological quantization presented in \cite{Ranada:1998vp}-\cite{Arrayas:2012eja}. In the last decade, the Ra\~{n}ada solutions were generalized in two ways. The first generalization is from knot fields to torus fields and was given in the works of Arrayas and Hoyos \cite{Arrayas:2011ia}-\cite{Hoyos:2015bxa}. The second generalization is from the electromagnetic fields in vacuum to electromagnetic fields in matter. Several authors have showed that topological fields can be found in various areas of physics such as: fluid physics \cite{Alves:2017ggb,Alves:2017zjt}, atmospheric physics \cite{Ranada:1996}, liquid crystals \cite{Irvine:2014}, plasma physics \cite{Smiet:2015}, optics \cite{Ren:2008zzf,deKlerk:2017qvq}, and superconductivity \cite{Trueba:2008sc}. (For a recent review of the knot solutions and their applications see \cite{Arrayas:2017sfq} and the references therein). More recently, new and important generalizations of the topological electromagnetic fields have been made to the non-linear electrodynamics, the fluid physics \cite{Goulart:2016orx}-\cite{Alves:2017zjt} as well as to the gravitational physics \cite{Kopinski:2017nvp} - \cite{Alves:2018wku}. A new method to obtain the electromagnetic knot solutions in Minkowski space, by completely solving the Maxwell equations first in de Sitter space and then conformally mapping to Minkowski space was presented in \cite{Lechtenfeld:2017tif,Kumar:2020xjr}. 

Due to their wide range of applications in both physics and mathematics, the topological electromagnetic fields represent an important field of science and an active line of research. 

In the present chapter, we are going to revisit the construction of the knot solutions of Maxwell's equations in Ran\~{a}da's approach. Also, we are going to briefly survey the different mathematical formulations of Maxwell's equations and to present the construction of the Hopf maps in electrodynamics. Of utmost importance for the understanding of the knot solutions is the factorization method of the 2-forms introduced by Bateman. We will review the application of Bateman's method to the topological electromagnetic fields. Finally, we are going to present the argument from \cite{Alves:2017ggb,Alves:2017zjt} where it was showed that knot solutions can also be found in the non-linear Born-Infeld electrodynamics and other non-linear generalizations of Maxwell's electrodynamics. We will focus our presentation on the electromagnetic fields in flat space-time. The generalization of the field line solutions to the gravitating electromagnetic fields is discussed in a different chapter of this volume \cite{Vancea:2019yjt}. The results reviewed here can be found in the original papers cited in the text. Other reviews are available too, most notably \cite{Arrayas:2017sfq} to which we refer for an updated list of references. In the Appendix, we collect some basic mathematical properties of the Hopf mapping. We adopt throughout this chapter the natural units in which $c = 1$.

\section{Maxwell's equations}  

In this section, we will briefly review the formulation of Maxwell's electrodynamics in terms of differential forms in the three-dimensions Euclidian space and in the four-dimensional space-time (the covariant formulation). This is a well-known material which can be found in standard textbooks on classical electrodynamics such as \cite{Jackson:1998nia,Frankel:1997ec}.

\subsection{Maxwell's equations in the three-dimensional formulation}

The two formulations of Maxwell's equations that are our concern in this paper are: the formulation in terms of differential forms and the covariant formulation. Both formulations are  equivalent to the one in terms of three-dimensional vectors that is known from the undergraduate textbooks on electrodynamics, see e. g. \cite{Jackson:1998nia}.

Consider the following standard form of Maxwell's equations in the Heaviside units
\begin{align}
\nabla \cdot \mathbf{D} & = \rho \, ,
\label{Maxwell-Gauss-El-1}
\\
\nabla \cdot \mathbf{B} & = 0 \, ,
\label{Maxwell-Gauss-Mag-2}
\\
\nabla \times \mathbf{E} & = 
- \frac{\partial \mathbf{B}}{\partial t}
 \, ,
\label{Maxwell-Faraday-3}
\\
\nabla \times \mathbf{H} & = 
\mathbf{J} + \frac{\partial \mathbf{D}}{\partial t}
\, .
\label{Maxwell-Ampere}
\end{align}
Here, $\mathbf{E}$ and $\mathbf{H}$ are the electric and magnetic intensities and 
$\mathbf{D}$ and $\mathbf{B}$ are the electric and magnetic flux densities, respectively. The sources of the electric and magnetic fields are the density of charge $\rho$ and the density of vector current $\mathbf{J}$. All vectors are three-dimensional and they are defined on the entire $\mathbb{R}^3$ at each instant of time if no other conditions are imposed on the system.

The set of equations (\refeq{Maxwell-Gauss-El-1}) - (\refeq{Maxwell-Ampere}) can be written in terms of differential forms from either $\Omega^{k}(\mathbb{R}^3 )$ or $\Omega^{k}(\mathbb{R}^{1,3} )$ which denote the differential forms of rank $k$ defined on the corresponding spaces. This abstract mathematical formulation is advantageous for at least two reasons - first, it provides a more economical formulation of the basic relations of the electromagnetism. Second, it provides the mathematical framework that is necessary for the generalization of the classical electromagnetism to curved space-times (see e. g. \cite{Frankel:1997ec}). The differential forms associated with the three-dimensional electromagnetic field are given in the Table
\refeq{tbl:fields}. 

\begin{table}[htbp]
  \begin{center}
   \label{table1}
   \caption{Field content of Maxwell's equations}
   \vspace{0.2cm}    
    \begin{tabular}{l|c|c|r}
      \textbf{Vector } & \textbf{Form} & \textbf{Form} & \textbf{Field}\\ 
      \textbf{notation} & \textbf{notation} & \textbf{rank} & \textbf{}\\ 
      \hline
      $\mathbf{E}$ & $\mathcal{E}$ & 1 - form & electric intensity\\ 
      $\mathbf{H}$ & $\mathcal{H}$ & 1 - form & magnetic intensity\\ 
      $\mathbf{D}$ & $\mathcal{D}$ & 2 - form & electric flux density\\ 
      $\mathbf{B}$ & $\mathcal{B}$ & 2 - form & magnetic flux density\\ 
      $\rho      $ & $\mathcal{Q}$ & 3 - form & charge density \\ 
      $\mathbf{J}$ & $\mathcal{J}$ & 2 - form & current density \\ 
    \end{tabular}
    \label{tbl:fields}  
  \end{center}
\end{table}

It is easy to show that the differential forms from the Table \refeq{tbl:fields} can be projected onto the orthonormal Cartesian basis of $\mathbb{R}^3$. The corresponding components are given by the following relations
\begin{align}
\mathcal{E} & = E_x dx + E_y dy + E_z dz \, ,
\nonumber
\\ 
\mathcal{H} & = H_x dx + H_y dy + H_z dz \, ,
\nonumber
\\
\mathcal{D} & = D_x dy \wedge dz + D_y dz \wedge dx + D_z dx \wedge dy \, ,
\nonumber
\\
\mathcal{B} & = B_x dy \wedge dz + B_y dz \wedge dx + B_z dx \wedge dy \, ,
\nonumber
\\
\mathcal{Q} & = \rho \, dx \wedge dy \wedge dz \, ,
\nonumber
\\
\mathcal{J} & = J_x dy \wedge dz + J_y dz \wedge dx + J_z dx \wedge dy \, .
\label{fields-components}
\end{align}
The basis $(dx, dy, dz)$ from the space $\Omega^1 (\mathbb{R}^3)$ of 1-forms is associated by the canonical procedure to the basis  $(x, y, z)$ from $\mathbb{R}^3$.  The second canonical basis  $(dx \wedge dy, dy \wedge dz, dz \wedge dx)$ in the space $\Omega^2 (\mathbb{R}^3)$ of 2-forms is constructed from the 1-forms $(dx, dy, dz)$ by taking their wedge product which is defined as follows
\begin{equation}
\wedge : \Omega^{k} \times \Omega^{s} \to \Omega^{k + s} \, ,
\qquad
(\omega , \sigma ) \to \omega \wedge \sigma \, .
\label{wedge-product}
\end{equation}
In the equation above, we have not specified which is the base space because the wedge product is defined in the same way on either $\mathbb{R}^3$ or $\mathbb{R}^{1,3}$. 

In order to write Maxwell's equations in terms of differential forms, one has to recall that the exterior derivative $d$ is defined as being the $\mathbb{R}$-linear map from $\Omega^k \to \Omega^{k+1}$ with the following properties
\begin{align}
d f & = \partial_{i} f \, d x^i \, ,
\label{exterior-derivative-1}
\\
d^2 \omega & = 0 \, , 
\label{exterior-derivative-2}
\\
d \left( \omega \wedge \sigma \right) & =
d \left( \omega  \right) \wedge \sigma
+ (-)^k \omega \wedge d \left( \sigma \right) \, .
\end{align}
Here, $f$ is an arbitrary smooth function and $\omega$ and $\sigma$ are arbitrary differential forms. Note that the exterior derivative is nilpotent, i. e. $d^2 = 0$.

The Hodge dual operation $\star$ is defined as the map from $\Omega^{n-k} \to \Omega^k $, where $n$ is the dimension of the base manifold, that satisfies the following relation 
\begin{equation}
\omega \wedge \left( \star \sigma \right) = \langle \omega , \sigma \rangle \mathrm{n} 
\, .
\label{Hodge-star}
\end{equation}
Here, $\mathrm{n}$ is an unitary vector and $\langle \cdot , \cdot \rangle$ is the scalar product. In particular, the action of the Hodge star operator on the components of an arbitrary $k$-form is given by the following equations \cite{Frankel:1997ec}
\begin{align}
\omega & = \frac {1}{k!} \omega_{i_{1},\dots ,i_{k}} 
d x^{i_{1}} \wedge \dots \wedge dx^{i_{k}}
= \sum _{i_{1}<\dots <i_{k}}
\omega_{i_{1},\dots ,i_{k}} 
dx^{i_{1}}\wedge \dots \wedge dx^{i_{k}} \, ,
\label{k-form}
\\
(\star \, \omega ) & = \frac {1}{(n-k)!} 
(\star \, \omega)_{i_{k+1},\dots , i_{n}} 
dx^{i_{k+1}} \wedge \dots \wedge dx^{i_{n}} \, .
\label{Hodge-k-form}
\end{align}
The above equations can be easily particularized to the three-dimensional case. For example, the Hodge duals to the elements of the canonical basis in the space of 1-forms are given by the following relations
\begin{equation}
\star d x  = d y \, d z \, ,
\qquad
\star d y  = d z \, d x \, ,
\qquad
\star d z  = d x \, d y \, .
\label{Hodge-3d-1-forms}
\end{equation}
Similarly, the Hodge duals of the elements of the  canonical basis in the space of 1-forms are obtained from the equation (\refeq{Hodge-k-form}) and they are given by the following relations
\begin{equation}
\star \, d y \, d z = d x  \, ,
\qquad
\star \, d z \, d x  = d y  \, ,
\qquad
\star \, d x \, d y = d z   \, .
\label{Hodge-3d-2-forms}
\end{equation}
The equations (\refeq{Hodge-3d-1-forms}) and (\refeq{Hodge-3d-2-forms})
illustrate the more general property of involution $\star \, (\star \, \omega ) = \omega$ in the three-dimensional Euclidean space. By using the equations (\refeq{Hodge-k-form}), (\refeq{Hodge-3d-1-forms}) and (\refeq{Hodge-3d-2-forms}), one can show that the exterior derivative $d$ can be decomposed in the canonical basis as follows
\begin{equation}
d = \left( \partial_x d x +  
\partial_y d y +
\partial_z d z
\right)
\wedge \, .
\label{exterior-derivative-3d}
\end{equation}
By using the exterior derivative from the equation (\refeq{exterior-derivative-3d}), one can write the Maxwell equations in terms of the three-dimensional differential forms. The result is given by the following set of equations 
\begin{align}
d \mathcal{D} & = \mathcal{Q} \, ,
\label{Maxwell-Gauss-el-3d}
\\
d \mathcal{B} & = 0 \, ,
\label{Maxwell-Gauss-mag-3d}
\\
d \mathcal{E} & = - \partial_t \mathcal{B} \, ,
\label{Maxwell-Faraday-3d}
\\
d \mathcal{H} & = \mathcal{J} + \partial_t \mathcal{D} \, .
\label{Maxwell-Ampere-3d}
\end{align}
It is easy to recognize in the equations (\refeq{Maxwell-Gauss-el-3d}) - (\refeq{Maxwell-Ampere-3d}) the familiar laws of the electromagnetism. They allow one to write the electromagnetic field in terms of potentials $\Phi$ and $\mathcal{A}$. That is made possible by the \emph{Poincaré's theorem}, that states that on a contractible manifold all closed forms ($d \omega = 0 $) are exact, i. e. for any exact form $\omega$ there exists a form $\sigma$ such that $\omega = d \sigma$ \cite{Frankel:1997ec}. By applying this mathematical result, one can show that the electric and magnetic 1-forms can be written as follows
\begin{align}
\mathcal{E} & = - d \Phi - \partial_t \mathcal{A} \, ,
\label{potentials-1}
\\
\mathcal{B} & = d \mathcal{A} \, .
\label{potentials-2}
\end{align}
It is an simple exercise to show that the equations (\refeq{Maxwell-Gauss-el-3d}) - (\refeq{Maxwell-Ampere-3d}) are invariant under the following gauge transformations
\begin{align}
\Phi & \rightarrow \Phi' = \Phi - \partial_t \Lambda \, ,
\label{gauge-1}
\\
\mathcal{A} & \rightarrow \mathcal{A}' = \mathcal{A} + d \Lambda \, ,
\label{gauge-2}
\end{align}
where $\Lambda$ is an arbitrary smooth function that plays the role of the gauge parameter.

The equations (\refeq{Maxwell-Gauss-el-3d}) - (\refeq{Maxwell-Ampere-3d}) describe the dynamics of the electromagnetic field alone. Their solutions are given in terms of charge and current densities, respectively, which are non-dynamical objects. As usual, the change of the source state is given by Newton's second law written for the Lorentz force, which has the following form
\begin{equation}
\mathcal{F}_{L} = \rho \mathcal{E} - \iota_{(\star \mathcal{J})} \mathcal{B} \, .
\label{Lorentz-force-3d}
\end{equation}
The interior product $\iota$ is defined as the contraction between a differential form from the space $\Omega^k$ and a vector field $X$. The interior product lowers the form degree by one, and its image belongs to the space $\Omega^{k-1}$. For example, if $\omega \in \Omega^2$ is a 2-form, its interior product with the vector field $X$ is the 1-form $\iota_{X} \omega$ for which the following equality holds 
\begin{equation}
\iota_{X} \omega (Y) = \omega (X, Y) \, ,  
\label{interior-product}
\end{equation}
where $Y$ is an arbitrary smooth vector field. The equation (\refeq{interior-product}) is the last equation needed to represent the dynamics of the electromagnetic field and its sources in terms of differential forms. Since in the rest of this chapter we will investigate the topological solutions of the electromagnetic field in the vacuum, that is away from its sources, we will ignore the Lorentz force as well as the charge and current densities.

\subsection{Maxwell equations in covariant formulation}

The geometric properties of the electromagnetic field are highlighted in the formulation of Maxwell's equations in terms of differential forms on $\mathbb{R}^3$. However, the fundamental symmetry of Maxwell's equations, which is the invariance under the Lorentz transformations, is explicitly displayed only in the Minkowski space-time $\mathbb{R}^{1,3}$. Therefore, it is important to write the Maxwell equations in terms of differential forms on $\mathbb{R}^{1,3}$. To this end, the canonical basis $\{ dx^i \}$ on $\mathbb{R}^3$ must be  extended to the corresponding basis on $\mathbb{R}^{1,3}$, denoted by $\{ d x^{\mu} \} = \{ dx^0 = dt, dx^i \}$, by including the time-like 1-form $dx^0$. Here, we are using the indices $\mu , \nu = 0, 1, 2, 3$ to denote the geometrical objects and their components in the Minkowski space-time. The electromagnetic 2-form field $F$ is defined by the following relation
\begin{equation}
F = B  + E \wedge d x^0 \, .
\label{EM-2-form}
\end{equation}
In this notation, the components of $F$ in the basis $\{ d x^{\mu} \wedge d x^{\nu} \}$ are the same as the components of the electromagnetic rank-2 antisymmetric tensor $F_{\mu \nu}$, namely
\begin{equation}
F = \frac{1}{2} F_{\mu \nu} d x^{\mu} \wedge d x^{\nu} \, .
\label{EM-tensor}
\end{equation}
The source of the electromagnetic field can also be written in terms of differential forms in four dimensions and it is expressed by the 1-form $J= J_{\mu} dx^{\mu}$ whose components are equal to the projections of the four-current $J^{\mu} = (\rho , \mathbf{J} )$ onto the space-time directions. By using these mathematical objects, one can write down the Maxwell equations in their most compact form as follows
\begin{align}
d F & = 0 \, ,
\label{Maxwell-4d-1}
\\
\star \, d \star F & = J \, .
\label{Maxwell-4d-2}
\end{align}
It is easy to show that the homogeneous equation (\refeq{Maxwell-4d-1}) corresponds to the magnetic Gauss law and to the Faraday law combined into one equation, while the inhomogeneous equation (\refeq{Maxwell-4d-2}) is the same as the electric Gauss law and the Maxwell-Amp\'{e}re law packed in to a single mathematical relation. 

The equations (\refeq{Maxwell-4d-1}) and (\refeq{Maxwell-4d-2}) fix to some extent the geometrical properties of the electromagnetic 2-form. Indeed, from the geometrical point of view, the field $F$ is a closed 2-form due to the equation (\refeq{Maxwell-4d-1}). From this property and from the nilpotency of the exterior derivative, i. e. $d^2 = 0$, we can derive the action of the differential form on the electromagnetic field which is given by the following equation
\begin{equation}
d F = d B + d E \wedge d x^0 \, .
\label{Maxwell-4d-1-a}
\end{equation}
Recall that the exterior derivative on $\mathbb{R}^{1,3}$ can be decomposed in to two terms: the exterior derivative $\mathbf{d}$ on the spatial subspace $\mathbb{R}^3 \in \mathbb{R}^{1,3}$ and the 1-form $dx^0$ along the time-like direction $\mathbb{R}$. This decomposition is given by the following equation 
\begin{equation}
d = dx^0 \wedge \partial_0 + \mathbf{d} \, . 
\label{ext-der-decomposed}
\end{equation} 
By plugging the equations (\refeq{ext-der-decomposed}) and (\refeq{Maxwell-4d-1-a}) into the equation (\refeq{Maxwell-4d-1}), one obtains the following set of equations
\begin{align}
\mathbf{d} E + \partial_0 B & = 0 \, ,
\label{Maxwell-4d-1-E}
\\
\mathbf{d} B & = 0 \, .
\label{Maxwell-4d-1-B}
\end{align}
The equations (\refeq{Maxwell-4d-1-E}) and (\refeq{Maxwell-4d-1-B}) are the homogeneous Maxwell equations written in terms of 1-forms on the Euclidean space $\mathbb{R}^3$. 

The three-dimensional inhomogeneous equations can be derived from the four-dimensional system, too. 
The outline of the derivation is the following. Firstly, note that the electromagnetic Hodge dual form $\star F$ can be obtained from $F$ by making the following replacements
\begin{equation}
E_j \to - B_j \, ,
\qquad
B_j \to E_j \, .
\label{F-dual-replacements}
\end{equation}
Secondly, decompose the form $\star F$ as follows
\begin{equation}
\star F = \boldsymbol{\star} E - \boldsymbol{\star} B \wedge d x^0 \, ,
\label{F-dual-decompose}
\end{equation}
where $\boldsymbol{\star}$ denotes the three-dimensional Hodge star operator. Thirdly, calculate the sequence of operations from the equation (\refeq{Maxwell-4d-2}). Then one obtains the following set of inhomogeneous equations
\begin{align}
\boldsymbol{\star} \, \mathbf{d} \boldsymbol{\star} E & = \rho \, ,
\label{Maxwell-4d-2-E}
\\
\boldsymbol{\star} \, \mathbf{d} \boldsymbol{\star} B 
- \partial_0 E 
& = \mathcal{J} \, .
\label{Maxwell-4d-2-B}
\end{align}
This concludes the derivation of the three-dimensional Maxwell equations in terms of differential forms from the four dimensional formulation. 

Let us make some observations about the electromagnetic fields away from their sources. In this case, the fields propagate in the vacuum. Their dynamics is still given by the equations (\refeq{Maxwell-4d-1}) and (\refeq{Maxwell-4d-2}), but with the supplementary condition $J = 0$. Due to this last relation, the symmetries of Maxwell's equations are enhanced to the group $SO(1,3) \times U(1) \times \mathcal{D}$, where $\mathcal{D}$ denotes the electromagnetic duality symmetry defined by the following transformations 
\begin{equation}
F \leftrightarrow \star F \, .
\label{duality-symmetry}
\end{equation}
The relation (\refeq{duality-symmetry}) implies that the 2-form $F$ can be written as a sum between a self-dual form $F_{+}$ and an anti self-dual form $F_{-}$. The corresponding equations are 
\begin{equation}
F = F_{+} + F_{-} \, ,
\qquad
\star F_{\pm} = \pm i F_{\pm} \, .
\label{self-dual-anti-self-dual}
\end{equation}
The last of the two equations from above is the result of the self-duality relation for 
2-forms in the Minkwoski space-time: $\star \star \, \omega = - \omega$. 

We end this section by observing that the formulation of the classical electrodynamics in terms of differential forms is completely equivalent with the formulation in terms of three-dimensional vector fields. However, each framework has its own advantages. The vector approach is useful for the visualization of the spatial distribution of fields and sources. The differential forms are interesting because they provide more economical equations, a deeper view of the symmetries and more information about the geometrical and topological properties of the electromagnetic fields.

\section{Knots in Maxwell's electrodynamics}

In this section, we are going to review the field line solutions of Maxwell's equations and their relationship with the Hopf knots. These solutions were discovered by Ra\~{n}ada and were communicated in \cite{Ranada:1989wc,Ranada:1990}. An important study of the Hopf fibration in the context of the classical electromagnetism can be found in the early pioneering work by Trautman \cite{Trautman:1977im}. In our presentation, we follow the very good indepth review of these solutions and their applications given in the reference \cite{Arrayas:2017sfq}. 

\subsection{Ran\~{a}da solutions}

The Ran\~{a}da solutions form a particular class of \emph{field line solutions} whose main feature is that the field lines completly characterize the electromagnetic field. In general, one can associate a field line to any smooth vector field through its integral flow. The path formed by the field line is tangent in each of its points to the vector obtained by taking the value of the vector field at that point. The picture of all field lines at a given instant of time can be used to visualize the state of the vector field at that instant.

As we can see from the equations (\refeq{Maxwell-4d-1}) and (\refeq{Maxwell-4d-2}), the properties of the sources determine the physical and geometrical properties of the electromagnetic field. In particular, the field lines of the electromagnetic field can have a very complex structure for non-trivial sources and can take a simple form in the vacuum.

In order to construct an electromagnetic field in terms of its field lines, we need a formal description of the latter. That can be given in terms of two smooth scalar complex fields on $\mathbb{R}^3$, namely
\begin{equation}
\phi : \mathbb{R}^3 \rightarrow \mathbb{C} \, , 
\qquad
\theta : \mathbb{R}^3 \rightarrow \mathbb{C} \, . 
\label{Ranada-complex-fields}
\end{equation}
By introducing the above functions, one can interpret the electric and the magnetic field lines as level lines of $\theta$ and $\phi$. In the covariant formulation, the corresponding electromagnetic fields take the following form \cite{Arrayas:2017sfq}
\begin{align}
F_{\mu \nu} & = g(\bar{\phi} , \phi ) 
\left( 
\partial_{\mu} \bar{\phi} \, \partial_{\nu} \phi
-
\partial_{\nu} \bar{\phi} \, \partial_{\mu} \phi
\right) \, ,
\label{Ranada-magnetic-solution}
\\
\star F_{\mu \nu} & = f(\bar{\theta} , \theta ) 
\left( 
\partial_{\mu} \bar{\theta} \, \partial_{\nu} \theta
-
\partial_{\nu} \bar{\theta} \, \partial_{\mu} \theta
\right) \, ,
\label{Ranada-electric-solution}
\end{align}
where $g$ and $f$ are smooth functions on $\theta$ and $\phi$, and the dual electromagnetic field has the following form
\begin{equation}
\star F_{\mu \nu} = \frac{1}{2} \epsilon_{\mu \nu \rho \sigma} F^{\rho \sigma} \, .
\label{dual-em}
\end{equation}
The components of the electromagnetic tensor $F_{\mu \nu}$ contain the electric and magnetic vector fields $\mathbf{E}$ and $\mathbf{B}$. These can be obtained from the following relations
\begin{align}
F & = \frac{1}{2} F_{ \mu \nu} dx^{\mu} \wedge dx^{\nu} = 
-\varepsilon_{j k l} B_{j} dx^{k} \wedge dx^{l} 
+ E_{j} dx^{j} \wedge d x^{0} \, , 
\label{F-field-components}
\\
\star F & = \frac{1}{2} \star F_{ \mu \nu} dx^{\mu} \wedge dx^{\nu} =
\varepsilon_{j k l} E_{j} dx^{k} \wedge dx^{l} 
+ B_{j} dx^{j} \wedge d x^{0} \, . 
\label{star-F-field-components}
\end{align} 
Note that the field line solutions (\refeq{Ranada-magnetic-solution}) and (\refeq{Ranada-electric-solution}) are given in the covariant formulation. Although they are explicitly Lorentz covariant and electromagnetically dual to each other, it is almost impossible to visualize the geometry of the electromagnetic field from the equations (\refeq{Ranada-magnetic-solution}) and (\refeq{Ranada-electric-solution}). This problem can be solved by invoking the equations (\refeq{F-field-components}) and (\refeq{star-F-field-components}) 
and by giving concrete values to the arbitrary functions $g$ and $f$. The first solution with the knot topology obtained in this way is the Ran\~{a}da solution
from \cite{Ranada:1989wc,Ranada:1990}. It has the following electric and magnetic decomposition
\begin{align}
E_{j} & = 
\frac{\sqrt{a} }{2 \pi i} \left( 1 + |\theta|^2 \right)^{-2} 
\varepsilon_{jkl} \, \partial_{k} \bar{\theta} \, \partial_{l} \theta \, ,
\label{Ranada-E}
\\
B_{j} & = 
\frac{\sqrt{a}}{2 \pi i} \left( 1 + |\phi|^2 \right)^{-2} 
\varepsilon_{jkl} \, \partial_{k} \bar{\phi} \, \partial_{l} \phi \, .
\label{Ranada-B}
\end{align}
The electromagnetic duality imposes some constraints on the fields $\theta$ and $\phi$. These constraints can be easily found by substituting the functions $g(\bar{\phi},\phi )$ and $f(\bar{\theta}, \theta)$ from the equations (\refeq{Ranada-E}) and (\refeq{Ranada-B}) into the equation (\refeq{duality-symmetry}), which leads to the following equations
\cite{Arrayas:2017sfq}
\begin{align}
\left( 1 + |\phi|^2 \right)^{-2} \varepsilon_{jmn} 
\partial_m \phi \partial_n \bar{\phi}
& = \left( 1 + |\theta|^2 \right)^{-2} 
\left(
\partial_0 \bar{\theta} \partial_j \theta
-
\partial_0 \theta \partial_j \bar{\theta}
\right)
\, ,
\label{fi-teta-1}
\\
\left( 1 + |\theta|^2 \right)^{-2} \varepsilon_{jmn} 
\partial_m \bar{\theta} \partial_n \theta
& = \left( 1 + |\phi|^2 \right)^{-2}
\left(
\partial_0 \bar{\phi} \partial_j \phi
-
\partial_0 \phi \partial_j \bar{\phi}
\right)
\, .
\label{fi-teta-2}
\end{align}
The equations (\refeq{fi-teta-1}) and (\refeq{fi-teta-2}) form a set of independent non-linear partial differential equations. The electric and magnetic field lines are the level curves of the solutions of the equations (\refeq{fi-teta-1}) and (\refeq{fi-teta-2}).

As one can see from the equations (\refeq{Ranada-E}) and (\refeq{Ranada-B}), the Ran\~{a}da fields correspond to particular functions $g$ and $f$. Recall that the \emph{null field} solutions of Maxwell's equations in vacuum are defined by the following equations
\begin{align}
E_j B_k \delta_{j k} = 0 \, ,
\label{null-field-1}
\\
\delta_{j k} \left( E^j E^k - B^j B^k \right) = 0 \, .
\label{null-field-2}
\end{align}
Then it is easy to verify that Ran\~{a}da fields $\mathbf{E}$ and $\mathbf{B}$ satisfy the first null field equation (\refeq{null-field-1}), but do not satisfy the second one. From that, we can conclude that the electric and magnetic fields are orthogonal to each other and so are their field lines. 

The parametrization of the electromagnetic 2-forms $F$ and $\star F$ in terms of $\phi$ and $\theta$ is not unique according to the equations (\refeq{Ranada-E}) and (\refeq{Ranada-B}). Other possibility is to use the so called \emph{Clebsch parametrization}. The main mathematical tool to construct it is the Darboux theorem \cite{Frankel:1997ec} that is satisfied by both $F$ and $\star F$. By using the Darboux theorem, we can write $F$ and $\star F$ in terms of four canonical 1-forms $d \sigma^a$, $ d \xi^a $, $d \rho^a$ and $d \zeta^a$
\cite{Arrayas:2017sfq} as follows
\begin{align}
F & = \delta_{ab} \, d \sigma^a \wedge d \xi^b \, ,
\, \, 
\label{Clebsch-F}
\\
\star F & = \delta_{ab} \, d \rho^a \wedge d \zeta^b \, ,
\, \, 
\label{Clebsch-star-F}
\\
d \sigma^a \wedge d \xi^a & = d \rho^a \wedge d \zeta^a = - i \tau_2 \, ,
\label{symplectic-form}
\end{align}
where the indices $a, b = 1, 2$ enumerate the 1-forms and $\tau_2$ is the Pauli matrix. The Clebsch representation of the electromagnetic field takes a simpler form for fields that satisfy the equation (\refeq{null-field-1}), namely
\begin{align}
F & = d \sigma \wedge d \xi\, ,
\, \, 
\label{Clebsch-F-1}
\\
\star F & = d \rho \wedge d \zeta \, .
\, \, 
\label{Clebsch-star-F-1}
\end{align}
Another important parametrization of the electromagnetic field is given in terms of 
\emph{Euler potentials} which are smooth scalar real fields on the Minkowski space-time \cite{Arrayas:2017sfq}
\begin{equation}
\alpha_a : \mathbb{R}^3 \rightarrow \mathbb{C} \, , 
\qquad
\beta_a : \mathbb{R}^3 \rightarrow \mathbb{C} \, , 
\label{Ranada-complex-fields-1}
\end{equation}
where $a = 1, 2$. The electric and magnetic fields have the following form
\begin{align}
E_{j} & = 
\varepsilon_{jkl} \, \partial_{k} \beta_2 \, \partial_{l} \beta_1 \, ,
\label{Euler-E}
\\
B_{j} & = 
\varepsilon_{jkl} \, \partial_{k} \alpha_2 \, \partial_{l} \alpha_1 \, .
\label{Euler-B}
\end{align}
By comparying the two sets of equations (\refeq{Ranada-E})-(\refeq{Ranada-B}) and (\refeq{Euler-E})-(\refeq{Euler-B}) with each other, one can easily find the following relation among the parameters of the Ran\~{a}da and Euler parametrizations, respectively,
\begin{align}
\beta_1 & = \left( 1+ |\theta|^2 \right)^{-1} \, ,
\qquad
\beta_2  = \frac{1}{2 \pi} \arg (\theta) \, ,
\label{beta-theta}
\\
\alpha_1 & = \left( 1+ |\phi|^2 \right)^{-1} \, ,
\qquad
\alpha_2  = \frac{1}{2 \pi} \arg (\phi) \, .
\label{alpha-beta}
\end{align} 
These representations are useful to express different properties of the field line solutions. Other representations can be found for more general topological electromagnetic fields. But before discussing this case, let us take a closer look at the Ran\~{a}da fields.

\subsection{Electromagnetic knot fields}

As we have seen above, the electromagnetic field characterized by the equations (\refeq{Ranada-magnetic-solution}) and (\refeq{Ranada-electric-solution}) is parametrized by $g$ and $f$. In particular, by choosing these functions as in the equations (\refeq{Ranada-E}) and (\refeq{Ranada-B}), the field line solutions display non-trivial topological properties. In order to see that, we note that $\phi$ and $\theta$ should be chosen such that the electromagnetic field and the observables constructed from it, e. g. energy, linear momentum, and angular momentum, be finite. Therefore, if the field is defined in the full $\mathbb{R}^3$, the regularity conditions to be imposed on the complex functions are the following 
\begin{equation}
|\phi(x)| \to 0 \, 
\,
\mbox{and}
\, \, 
|\theta(x)| \to 0 \, \,
\mbox{if}
\, \, 
|\mathbf{x}| \to \infty \, .
\label{regularity}
\end{equation}
The boundary conditions (\refeq{regularity}) imply that the defining domain of $\phi$ and $\theta$ is the compactification $\mathbb{R}^3 \cup \{ \infty \} = S^3$ at any given value of $t$. Since $\phi$ and $\theta$ take values in the compactification $\mathbb{C} \cup \{ \infty \} = S^2$, they can be viewed as two families of one-parameter maps $\phi(x) = \{ \phi_t (\mathbf{x}) \}_{t \in \mathbb{R}}$ and $\theta(x) = 
\{ \theta_t (\mathbf{x}) \}_{t \in \mathbb{R}}$ from $S^3$ to $S^2$. Here, it is convenient to use the vector notation to discuss separately the electric and the magnetic field lines, respectively. 

The fields $\mathbf{E}$ and $\mathbf{B}$ can be written in terms of electromagnetic potentials $\mathbf{C}$ and $\mathbf{A}$ in a symmetric way
\begin{equation}
E_j = \varepsilon_{jkl} \partial_k C_l \, ,
\qquad
B_j = \varepsilon_{jkl} \partial_k A_l \, .
\label{C-A-potentials}
\end{equation}
Note that $\mathbf{C}$ is dependent on $\mathbf{A}$, as they are related by the following equation
\begin{equation}
\varepsilon_{jmn} \partial_m C_n = - \partial_0 A_j \, .
\label{C-A-relation}
\end{equation}
However, it is convenient to keep $\mathbf{C}$ explicit, as it makes the electric-magnetic duality more symmetric. The Chern-Simons integrals associated to the electric and magnetic fields and their potentials can be used to define the helicities of the electromagnetic field as follows
\begin{align}
H_{ee} & = \int d^3 x \, \delta_{ij} E_i C_j = 
\int d^3 x \, \varepsilon_{jkl} C_j \partial_k C_l \, , 
\label{helicity-Hee}
\\
H_{mm} & = \int d^3 x \, \delta_{ij} B_i A_j = 
\int d^3 x \, \varepsilon_{jkl} A_j \partial_k A_l \, , 
\label{helicity-Hmm}
\\
H_{em} & = \int d^3 x \, \delta_{ij} B_i C_j = 
\int d^3 x \, \varepsilon_{jkl} C_j \partial_k A_l \, , 
\label{helicity-Hem}
\\
H_{mm} & = \int d^3 x \, \delta_{ij} E_i A_j = 
\int d^3 x \, \varepsilon_{jkl} A_j \partial_k C_l \, . 
\label{helicity-Hme}
\end{align}
The pure electric and magnetic helicites given above are the Hopf indexes of the corresponding electric and magnetic field lines (see the Appendix). If one substitutes the equations (\refeq{C-A-potentials}) into Maxwell's equations (\refeq{Maxwell-Gauss-El-1}) - (\refeq{Maxwell-Ampere}) in the vacuum, one obtains the following set of equations
\begin{align}
\varepsilon_{jmn} \partial_m 
\left( 
\frac{\partial A_{n}}{\partial x^0}
+ \varepsilon_{nrs} \partial_r C_s
 \right) & = 0 \, ,
\label{Maxwel-CA-1}
\\
\varepsilon_{jmn} \partial_m 
\left( 
\frac{\partial C_{n}}{\partial x^0}
- \varepsilon_{nrs} \partial_r A_s
 \right) & = 0 \, .
\label{Maxwel-CA-2}
\end{align}
From the vector calculus, we conclude that there are two scalar functions $\kappa_1$ and $\kappa_2$ such that
\begin{align}
\frac{\partial A_{j}}{\partial x^0}
+ \varepsilon_{jmn} \partial_m C_n
& = \partial_j \kappa_1 \, ,
\label{Maxwel-CA-1-sol}
\\
\frac{\partial C_{j}}{\partial x^0}
- \varepsilon_{jmn} \partial_m A_n
& = \partial_j \kappa_2 \, .
\label{Maxwel-CA-2-sol}
\end{align}
By using the equations (\refeq{Maxwel-CA-1}) - (\refeq{Maxwel-CA-2-sol}), the following equations are obtained
\begin{align}
\delta_{mn} \frac{\partial (A_m B_n)}{\partial x^0} 
+ 2 \delta_{mn} E_m B_n 
- \partial_k \left( \varepsilon_{krs} A_k E_s 
-\kappa_1 B_k
\right) & = 0 \, .
\label{identity-1}
\\
\delta_{mn} \frac{\partial (C_m E_n)}{\partial x^0} 
+ 2 \delta_{mn} E_m B_n 
+\partial_k \left( \varepsilon_{krs} C_k B_s 
+\kappa_2 E_k
\right) & = 0 \, .
\label{identity-2}
\end{align}
From these, it follows that
\begin{align}
\frac{\partial (H_{mm} - H_{ee} )}{\partial x^0} + 4 \int d^3 x
\delta_{mn} E_m B_n & = 0 \, ,
\label{H-conservation-1}
\\
\frac{\partial (H_{mm} + H_{ee} )}{\partial x^0} & = 0 \, .
\label{H-conservation-2}
\end{align}
The equations (\refeq{H-conservation-1}) and (\refeq{H-conservation-2}) show that the pure electric and magnetic helicities are conserved if the fields satisfy the first null field condition (\refeq{null-field-1}).

Let us quote here the explicit Hopfion solution obtained in \cite{Ranada:1997}. The fields $\phi$ and $\theta$ have the following form
\begin{align}
\phi & = \frac{(ax_1 - x_0 x_3) + i(ax_2 + x_0 (a-1)}{(ax_3 + x_0 x_1) + i[a(a-1) - x_0 x_1]}  \, ,
\label{hopfion-1-phi}
\\
\theta & = \frac{[ax_2 + x_0 (a-1)] + i(ax_3 + x_0 x_1)}{(ax_1 - x_0 x_3) + i[a(a-1) - x_0 x_2]}  \, , 
\label{hopfion-1-theta}
\end{align}
where $x_{\mu}$'s are dimensionless coordinates and 
\begin{equation}
a = \frac{r^2-x^{2}_{0} + 1}{2} \, ,
\qquad
r = \sqrt{\delta_{mn} x_{m}^{2} x_{n}^{2} } \, .
\label{hopfion-1-constants}
\end{equation}
The electric and magnetic fields derived from the scalars given by the equations (\refeq{hopfion-1-phi}) and (\refeq{hopfion-1-theta}) have the following form
\begin{align}
\mathbf{E} & = \frac{1}{\pi} \frac{q \mathbf{H}_1 - p \mathbf{H}_2}{(a^2 + x^{2}_{0} )^3} \, ,
\label{Hopfion-1-E}
\\
\mathbf{B} & = \frac{1}{\pi} \frac{q \mathbf{H}_1 + p \mathbf{H}_2}{(a^2 + x^{2}_{0} )^3} \, ,
\label{Hopfion-1-B}
\end{align}
where
\begin{align}
\mathbf{H}_1 & = 
\left( x_2 + x_0 - x_1 x_3 \right)\mathbf{e}_1
-\left[ x_1 + \left( x_2 + x_0 \right) x_3 \right] \mathbf{e}_2
\nonumber
\\
& + \frac{1}{2}
\left[
-1 - x^{2}_{3} + x^{2}_{1} + \left( x_2 + x_0 \right)^2
\right]\mathbf{e}_3
\label{Hopfion-1-H1}
\\
\mathbf{H}_2 & = + \frac{1}{2}
\left[
1 + x^{2}_{1} - x^{2}_{3} - \left( x_2 + x_0 \right)^2
\right]\mathbf{e}_1
+
\left[ - x_3 + \left( x_2 + x_0 \right) x_1 \right] \mathbf{e}_2
\nonumber
\\
& + \left( x_2 + x_0 + x_1 x_3 \right)\mathbf{e}_3 \, ,
\label{Hopfion-1-H2}
\end{align}
and
\begin{equation}
p = x_0 \left( x^{2}_{0} - 3 a^2 \right) \, ,
\qquad
q = a \left( a^2  - 3 x^{2}_{0} \right) \, .
\label{Hopfion-1-constants-1}
\end{equation}
Here, we have denoted by $\{ \mathbf{e}_l \}$ the unit vectors of the Cartesian basis in the spatial directions. The Hopf indices of the above Hopfion solution are
$H(\phi) = H(\theta) = 1$ \cite{Arrayas:2017sfq}. 

Another example of Hopfion is given by the Hopf map defined by the following scalar fields
\begin{align}
\phi & = \frac{2(x_1 + i x_2)}{2x^{2}_{3} + i (\delta_{mn}x^m x^n  - 1 )} \, ,
\label{Ranada-Hopf-1-phi}
\\
\theta & = \bar{\phi} \, .
\label{Ranada-Hopf-1-theta}
\end{align}
The electromagnetic field corresponding to these scalars has the following components
\begin{align}
E_m & = \frac{1}{\sqrt{(2 \pi)^3}} \int d^3 k 
\left[
P_m (k_j) \cos \left( \eta_{\mu \nu} k^{\mu} x^{\nu} \right)
-
Q_m (k_j) \sin \left( \eta_{\mu \nu} k^{\mu} x^{\nu} \right)
\right] \, ,
\label{Ranada-Hopf-1-E}
\\
B_m & = \frac{1}{\sqrt{(2 \pi)^3}} \int d^3 k 
\left[
P_m (k_j) \cos \left( \eta_{\mu \nu} k^{\mu} x^{\nu} \right)
+
Q_m (k_j) \sin \left( \eta_{\mu \nu} k^{\mu} x^{\nu} \right)
\right] \, ,
\label{Ranada-Hopf-1-B}
\end{align}
where
\begin{align}
\mathbf{P} & = 
\frac{e^{-k_0}}{\sqrt{2 \pi}}
\left(
- \frac{k_1 k_3}{k_0}
,
\frac{k_0 k_2 + k^{2}_{2} + k^{2}_{3} }{k_0}
,
- \frac{k_0 k_1 + k_1 k_2 }{k_0}
\right) \, ,
\label{Ranada-Hopf-1-P}
\\
\mathbf{Q} & = 
\frac{e^{-k_0}}{\sqrt{2 \pi}}
\left(
- \frac{k_0 k_2 + k^{2}_{1} + k^{2}_{2} }{k_0}
,
\frac{k_1 k_3}{k_0}
,
\frac{k_0 k_3 + k_2 k_3 }{k_0}
\right) \, .
\label{Ranada-Hopf-1-Q}
\end{align}
The fields given by the equations (\refeq{Ranada-Hopf-1-E}) and (\refeq{Ranada-Hopf-1-B}) describe a particular wave packet that travels along the $x_3$ axis and has the following energy, linear momentum, and angular momentum densities
\begin{equation}
\mathit{E} = 2 \, ,
\qquad \mathbf{p} = \left( 0, 0, 1 \right) \, ,
\qquad \mathbf{L} = \left( 0, 0, 1 \right) \, .
\label{Ranada-Hopf-1-E-p-L}
\end{equation} 
The exact values of $\mathit{E}$, $\mathbf{p}$ and $\mathbf{L}$ are calculated in a dimensionless parametrization of the space-time coordinates. From the equations (\refeq{Ranada-Hopf-1-E-p-L}), it follows that the mass of the wave packet is finite and it has the value $m^2 = 3$.

\section{Electric and magnetic knots in Bateman parametrization}

In the previous section, we have discussed several parametrizations of the field line solutions. In this section, we present the factorized parametrization of the self-dual electromagnetic fields in terms of 1-forms given by Bateman in \cite{Bateman:2016}. This representation is useful for the generalization of the field line solutions to the gravitating electromagnetic fields \cite{Vancea:2017tmx} as well as to the non-linear electromagnetism \cite{Alves:2017ggb,Alves:2017zjt}. For extensive reviews of the properties of the Hopfions in the Bateman representation see  \cite{Hoyos:2015bxa,Arrayas:2017sfq,Thompson:2014owa}.

Let us recall the covariant form of Maxwell's equation in the vacuum given by the equations (\refeq{Maxwell-4d-1}) and (\refeq{Maxwell-4d-2}), namely
\begin{align}
d F & = 0 \, ,
\label{Maxwell-4d-1-vacuum}
\\
d \star F & = 0 \, .
\label{Maxwell-4d-2-vacuum}
\end{align}
The equation (\refeq{Maxwell-4d-1-vacuum}) states that the electromagnetic 2-form field is closed. Then $F$ can be written in terms of two scalar complex functions
\begin{equation}
\alpha : \mathbb{R}^3 \rightarrow \mathbb{C} \, , 
\qquad
\beta : \mathbb{R}^3 \rightarrow \mathbb{C} \, ,
\label{Bateman-complex-fields}
\end{equation}
as follows
\begin{equation}
F = d\alpha \wedge d\beta \, .
\label{F-field-Bateman}
\end{equation}
It is relevant to write the above equation in terms of the electric and magnetic fields. To this end, we use the components of the electromagnetic 2-form $F$ which are the same as the components of the electromagnetic tensor $F_{\mu \nu}$. It follows from the equation (\refeq{F-field-Bateman}) that $F_{\mu \nu}$ satisfies the following relation
\begin{equation}
i \varepsilon^{\mu \nu \rho \sigma} F_{\rho \sigma}
- 2 F^{\mu \nu}  = 2 \varepsilon^{\mu \nu \rho \sigma} 
\partial_{\rho} \alpha \partial_{\sigma} \beta \, .
\label{F-field-covariant}
\end{equation}
By definition, the electric and magnetic fields are given by the following equations
\begin{equation}
- E^{m} = F^{0m} \, ,
\qquad
B^{m} = \frac{1}{2} \varepsilon^{mnp} F_{np} \, .
\label{F-E-B-components}
\end{equation}
Some algebraic manipulations of the equation (\refeq{F-field-covariant}) lead to the following relation between the complex scalars
\begin{equation}
B_m - i E_m 
=
 i 
\left(
\partial_0 \alpha \partial_m \beta 
- \partial_0 \beta \partial_m \alpha
\right) \, .
\label{alpha-beta-constraint-0}
\end{equation}
We know from the equation (\refeq{self-dual-anti-self-dual}) that the electromagnetic field in the vacuum is either self-dual or anti self-dual according to the eigenvalues of the $\star$ operation. That implies that the functions $\alpha$ and $\beta$ must obey the following constraint
\begin{equation}
\nabla \alpha \times \nabla \beta
=
\pm i 
\left(
\partial_0 \alpha \nabla \beta - \partial_0 \beta \nabla \alpha
\right) \, .
\label{alpha-beta-constraint}
\end{equation}
In order to understand the geometry of the electric and magnetic fields on $\mathbf{R}^3$, it is useful to introduce the Riemann-Silberstein vector
\begin{equation}
\mathbf{F} = \mathbf{B} \pm i \mathbf{E} \, , 
\label{F-vector-field-Bateman}
\end{equation}  
where $\mathbf{E}$ and $\mathbf{B}$ can be complex. Also, it is required that $\mathbf{F}$ be a solution of Bateman's equation
\begin{equation}
\delta_{mn} F_m F_n  = 0 \, .
\label{Bateman-equation}
\end{equation}
Note that in general the norm of the field $\mathbf{F}$ is non-zero
\begin{equation}
\delta_{mn} \bar{F}_m F_n  \neq 0 \, ,
\label{Bateman-equation-norm}
\end{equation}
where the bar stands for the complex conjugate. 
One can easily write the equations (\refeq{Bateman-equation}) and (\refeq{Bateman-equation-norm}) on components. The result is the following
\begin{align}
\delta_{mn} \left( B_m B_n -  E_m E_n \right)
\pm 2 i \delta_{mn} E_m B_n & = 0 \, , 
\label{Bateman-equation-1}
\\
\delta_{mn} \left( \bar{B}_m B_n +  \bar{E}_m E_n \right)
& = 0 \, .
\label{Bateman-equation-norm-1}
\end{align}
If the vector fields $\mathbf{E}$ and $\mathbf{B}$ are real, the left-hand side of the equation (\refeq{Bateman-equation-1}) is an invariant of the electromagnetic field and the left-hand side of the equation (\refeq{Bateman-equation-norm-1}) is the energy density of the electromagnetic field. The equations (\refeq{Bateman-equation-1}) and (\refeq{Bateman-equation-norm-1}) define the so called \emph{null fields} \cite{Kedia:2013bw}. 

It is easy to verify that the Bateman solutions conserve the energy, momentum and the angular momentum of the electromagnetic field. These conservation laws follow from the Lorentz symmetry. Also, the $U(1)$ symmetry of Maxwell's equations implies that there is a conserved four-current whose components are given by the following relations
\begin{align}
\rho & = \frac{1}{2} \delta_{mn}
\left(
E_m E_n + B_m B_n
\right) \, ,
\label{Bateman-conserved-rho}
\\
J_k & = \varepsilon_{kmn} E_m B_n \, .
\label{Bateman-conserved-J}
\end{align}
Beside the conserved quantities discussed above, there are new topologically conserved charges, the helicities from the equations (\refeq{helicity-Hee})-(\refeq{helicity-Hmm}).
The Bateman equations admit solution with knot as well as toric topologies. In the Bateman parametrization, different solutions can be related to each other, or derived from each other, due to the following important theorem \cite{deKlerk:2017qvq}:

{\it
Let $\alpha$ and $\beta$ be two smooth complex scalar fields on $M$ that satisfy the Bateman relation (\refeq{alpha-beta-constraint}). Then for any two arbitrary smooth complex functions $f$ and $g$ defined on $\mathbb{C}^2$, the following 2-form exists
\begin{equation}
\mathcal{F} : = d f(\alpha, \beta ) \wedge d g (\alpha , \beta ) \, ,
\label{Bateman-F-general}
\end{equation} 
and $\mathcal{F}$ has the following properties:
\begin{align}
d \mathcal{F} & = 0 \, ,
\label{Bateman-F-general-closed}
\\
\star \mathcal{F} & = \pm i \mathcal{F} \, .
\label{Bateman-F-general-self-dual} 
\end{align}}
There are several known solutions in the literature constructed with the Bateman method. Let us cite here the Hopfion obtained in \cite{Kedia:2013bw} whose complex functions are given by the following relations
\begin{align}
\alpha & = \frac{- x^{2}_{0} + \delta_{mn}x^m x^n - 1 + 2ix_3}{- x^{2}_{0} + \delta_{mn}x^m x^n +1 +2ix_0} \, ,
\label{Bateman-Hopfion-1-alpha}
\\
\beta & = \frac{x_1 - i x_2}{- x^{2}_{0} + \delta_{mn}x^m x^n +1 +2ix_0} \, .
\label{Bateman-Hopfion-1-beta}
\end{align}
The functions $\alpha$ and $\beta$ given above satisfy the following relation
\begin{equation}
|\alpha|^2 + |\beta|^2 = 1 \, .
\label{Bateman-alpha-beta-constraint}
\end{equation}
The electromagnetic vector $\mathbf{F}$ that is obtained from $\alpha$ and $\beta$ 
has the following form
\begin{align}
\mathbf{F} & =\frac{4}{(- x^{2}_{0} + \delta_{mn}x^m x^n + 1 + 2 i x_0)^3}
\nonumber
\\
& \times 
\left[
\begin{array}{c}
\left(x_0 - x_1 - x_3 + i \left(x_2-1 \right) \right) 
\left( x_0 + x_1 - x_3 - i \left( x_2 + 1 \right) \right) \\
-i \left(x_0 - x_2 - x_3 - i \left( x_1 + 1 \right) \right)
\left( x_0 + x_2 - x_3 + i \left( x_1-1 \right) \right) \\
2 \left(x_1 - i x_2 \right) \left( x_0 - x_3 -i \right)  \\
\end{array}
\right] \, ,
\label{Bateman-Hopfion-1-F}
\end{align}
where $\mathbf{F}$ is in a column vector notation. The general properties of the field (\refeq{Bateman-Hopfion-1-F}) are discussed in \cite{Kedia:2013bw,Hoyos:2015bxa}.

New Hopfions can be obtained from a given solution by performing infinitesimal conformal transformations, or a subgroup of them, on the functions $\alpha$ and $\beta$. In order to see that, recall that the scalar functions change under the infinitesimal coordinate trasformations as follows \cite{Hoyos:2015bxa}
\begin{align}
x^{\mu} & \rightarrow x^{'\mu} = x^{\mu} + \xi^{\mu}
\label{Bateman-deformations-coordinates}
\\
\alpha(x) & \rightarrow \alpha'(x') = \alpha (x) + \xi^{\nu} \partial_{\nu} \alpha (x) \, ,
\label{Bateman-deformations-alpha}
\\
\beta(x) & \rightarrow \beta'(x') = \beta (x) + \xi^{\nu} \partial_{\nu} \beta (x) \, ,
\label{Bateman-deformations-beta}
\end{align}
where $\xi = \xi^{\mu} \partial_{\mu}$ is an arbitrary infinitesimal smooth vector field on $\mathbb{R}^{1,3}$. The equation (\refeq{alpha-beta-constraint}) is invariant under the transformations (\refeq{Bateman-deformations-coordinates}) - (\refeq{Bateman-deformations-beta}) if the following condition is satisfied \cite{Hoyos:2015bxa}
\begin{equation}
\epsilon_{rmn}
\left[ 
\delta_{jr}
\left(
\partial_0 \xi_0 - \partial_s \xi_s
\right)
+ i
\epsilon_{jrs}
\left( 
- \partial_0 \xi_s + \partial_s \xi_0
\right)
+
\partial_j \xi_r + 
\partial_r \xi_j
\right]
\, 
\partial_m \alpha \partial_n \beta =
0 \, .
\label{Bateman-transformations-condition}
\end{equation}
The generators of the special conformal transformations are the vector fields 
\begin{equation}
\xi_{\mu} = a_{\mu} \eta_{\mu \nu} x^{\mu \nu} 
- 2 a_{\nu} x^{\nu} x^\mu \, .
\label{Bateman-special-conformal-transformations}
\end{equation}
It is easy to show that any vector field of the form (\refeq{Bateman-special-conformal-transformations}) is a solution of the equation (\refeq{Bateman-transformations-condition}). 
By using this property of the special conformal transformations, the authors of \cite{Hoyos:2015bxa} obtained new Bateman solutions characterized by integer powers $p$ and $q$ of the scalar functions $\alpha$ and $\beta$. The plane wave solution is one example of electromagnetic field from which new solutions can be generated by conformal deformations. The plane waves are characterized by the following scalar functions 
\begin{equation}
\alpha = e^{\imath (x_3-x_0)} \, ,
\qquad 
\beta = x_1 + i x_2 . 
\label{Bateman-Hopfion-plane-wave}
\end{equation}
The new functions that can be obtained from the equation (\refeq{Bateman-Hopfion-plane-wave}) by conformal deformations have the following form
\begin{align}
\alpha' & = \exp
\left[
-1 + 
\frac{i \left( x_0 + x_3 - i \right)}
{ 1 - x^{2}_{0} 
+ \delta_{mn} x^m x^n + 2 i x_0}
\right],
\label{Bateman-Hopfion-plane-wave-conformal-1}
\\
\beta' & = \frac{ x_1 + i x_2}{1 - x^{2}_{0} 
+ \delta_{mn} x^m x^n + 2 i x_0} 
\, .
\label{Bateman-Hopfion-plane-wave-conformal-2}
\end{align}
The electric and magnetic fields built from $\alpha'$ and $\beta'$ have \emph{toric topology}. A generalization of this method to more complex knotted electromagnetic fields was given in \cite{Kedia:2016nwk}. The conformal deformations method was applied to the study of the physical properties of the optical vorticies in \cite{Dennis:2010,deKlerk:2017qvq}.

\section{Knots in nonlinear electrodynamics}

A very important application of Bateman method to non-linear electrodynamical models that generalize Maxwell's electrodynamics in the strong field regime, was given recently in \cite{Alves:2017ggb,Alves:2017zjt}.
In these works, the authors showed that there are knot solutions of the equations of motion in any non-linear extension of Maxwell's electrodynamics that satisfies strong field requirement. 

Let us see how the Bateman method can be applied to the Born-Infeld electrodynamics. The non-linear Born-Infeld action has the following form
\begin{equation}
S_{BI} = - \gamma^2 \int d^4 x \left( \sqrt{1+ F - P^2} - 1 \right) \, ,
\label{NL-Born-Infeld-action}
\end{equation}
where $\gamma$ is a constant of dimension 2 and $L$ and $P$ are the usual Lorentz invariants defined as follows
\begin{align}
L & = \gamma^{-2} \delta_{m n}\left( B^m B^n - E^m E^n \right)
\, ,
\label{lorentz-invariant-L}
\\
P & = \gamma^{-2} \delta_{m n} E^m B^n \, .
\label{Lorentz-invariant-P}
\end{align}
The scalars $L$ and $P$ are zero for all null-fields that satisfy the equations (\refeq{null-field-1}) and (\refeq{null-field-2}). 

In order to prove that the theory described by the action $S_{BI}$ has Hopfion solutions, the following argument has been developed in \cite{Alves:2017ggb,Alves:2017zjt}. Starting from the action (\refeq{NL-Born-Infeld-action}), define new vector fields $\mathbf{H}$ and $\mathbf{D}$ whose components can be written in a notation analogous to Maxwell's electrodynamics
\begin{equation}
H_m = - \frac{\partial \mathcal{L}_{BI}}{\partial B^m} \, ,
\qquad
D_m = \frac{\partial \mathcal{L}_{BI}}{\partial E^m} \, .
\label{NL-BI-fields}
\end{equation}
Here, $\mathcal{L}_{BI}$ is the Lagrangian density from the action (\refeq{NL-Born-Infeld-action}). From it, one obtains the explicit form of $H_m$ and $D_m$ as follows 
\begin{align}
H_m & = - \frac{1}{\sqrt{1+ F - P^2}} 
\left(
B_m - P E_m
\right)
\, ,
\label{NL-BI-H}
\\
D_m & = 
- \frac{1}{\sqrt{1+ F - P^2}} 
\left(
E_m - P B_m
\right)
\, .
\label{NL-BI-D}
\end{align}
By applying the variational principle to the action (\refeq{NL-Born-Infeld-action}), the following equations of motion are obtained  
\begin{align}
\partial_m D_m & = 0 \, ,
\label{NL-PB-Eq1}
\\
\partial_m B_m & = 0 \, ,
\label{NL-PB-Eq2}
\\
\varepsilon_{m n r} \partial_n E_r & = - \partial_0 B_m \, ,
\label{NL-PB-Eq3}
\\
\varepsilon_{m n r} \partial_n H_r & = \partial_0 D_m \, .
\label{NL-PB-Eq4}
\end{align}
The above equations show that the null fields from the relations (\refeq{null-field-1}) and (\refeq{null-field-2}) belong to a theory in which the equations of motion are of Maxwellian type. Therefore, by using the results presented in Section 2, one concludes that there are null field Hopfion solutions in the non-linear Born-Infeld electrodynamics. 

The same argument can be generalized to all non-linear actions that in the weak field limit can be reduced to Maxwell's electrodynamics. Another interesting generalization is based on the observation that the equations of the classicall electrodynamics are similar to the equations of the fluid mechanics. Then by using the same reasoning as before, Hopfion solutions of the fluid flow lines were showed to exist in \cite{Alves:2017ggb,Alves:2017zjt}. 

\section*{Appendix - Hopf map}

In this Appendix, we briefly review the definition and some basic properties of the Hopf map that have been used in the text. A classical reference on this topic is the textbook \cite{Milnor:1997} in which Hopf's theory is presented from a geometrical point of view.

Consider a $n$-dimensional unit sphere $S^n$ defined as follows
\begin{equation}
S^n = \{ \, \mathbf{x} \in R^{n+1} : \,
\delta_{ij} x_i x_j = 1, \, i, j = 0, 1, \ldots , n \, \} \, .
\label{Sn-sphere}
\end{equation}
The \emph{Hopf fibration} is the mapping $h: S^3 \to S^2 \simeq \mathbb{CP}^1$ defined by the following \emph{Hopf map}
\begin{equation}
h(x_0, x_1, x_2, x_3 ) = 
\left(
x^{2}_{0} + x^{2}_{1} - x^{2}_{2} - x^{2}_{3} \, ,
2(x_0 x_3 + x_1 x_2) \, ,
2( x_1 x_2 - x_0 x_3 )
\right) \, .
\label{Hopf-map}
\end{equation}
Alternatively, the Hopf map can be written in terms of the following complex coordinates on the spheres $S^2$ and $S^3$
\begin{align}
S^2 & = \{ (x,z) \in \mathbb{R} \times \mathbb{C} \, : \, 
x^2 + |z|^2 = 1
\} \, ,
\label{S2-complex}
\\
S^3 & = \{ (z_1,z_2 ) \in \mathbb{C}^2 \, : \, |z_1|^2 + |z_2|^2 = 1 
\}
\, .
\label{S3-complex}
\end{align}
Then the Hopf map takes the following form 
\begin{equation}
h(z_1 , z_2) = 
\left(
2 z_1 \bar{z}_2
,
|z_1|^2 - |z_2|^2
\right) \, .
\label{Hopf-map-complex}
\end{equation}
Since the projective space is a quotient set $\mathbb{CP}^1 \simeq S^3/U(1)$, the action of $U(1)$ on $S^3$ defines the fibres over $S^3$. These are mapped into the fibres over $S^2$ as follows. Consider the equivalence class of points of $S^3$ defined by the following relation
\begin{equation}
(z'_{1}, z'_{2} ) \sim (z_1, z_2) \, : \, 
\mbox{if} \, \, \exists \, \lambda \in \mathbb{C} 
\, \, \mbox{such that} 
\, \,
(z'_{1}, z'_{2} ) = (\lambda z_1, \lambda z_2) \, .
\label{U(1)-equivalence}
\end{equation}
Then the Hopf map of $(z'_{1}, z'_{2} )$ satisfies the following relation
\begin{equation}
h(z'_{1}, z'_{2} ) = |\lambda|^2 h(z_1 , z_2) \, .
\label{Hopf-U(1)-equivalence}
\end{equation}
The points $(z'_{1}, z'_{2} )$  and $(z_1, z_2)$ of $S^3$ belong to a fibre over $S^2$ if they are mapped onto the same point of $S^2$. This represents a constraint on the parameter $\lambda$ which is satisfied by 
any $\lambda$ of the form $\lambda = \exp (i\vartheta)$, where $\vartheta \in [0, 2 \pi)$. Thus, $\lambda$ necessarily belongs to the defining representation of $U(1)$. 

From the geometrical point of view, the fibre of $S^3$ is the great circle that contains $(z_1, z_2)$. An useful parametrization of fibres is given by the following relations
\begin{equation}
z_1 = \exp \left( i\xi + \frac{\varphi}{2} \right) 
\sin \left( \frac{\vartheta}{2}\right) \, ,
\qquad
z_2 = \exp \left( i\xi - \frac{\varphi}{2} \right) 
\cos \left( \frac{\vartheta}{2}\right) \, .
\end{equation}
It is easy to see that the Hopf map $h$ takes the points from the fibres of $S^3$ to the following points on $S^2 \subset \mathbb{R}^3$
\begin{align}
x_1 & = 2 |z_1 | | z_2 | \cos(\varphi) \, ,
\label{Hopf-1}
\\
x_2 & = 2 |z_1 | | z_2 | \sin(\varphi) \, ,
\label{Hopf-2}
\\
x_3 & = |z_1 |^2 - | z_2 |^2 \, .
\label{Hopf-3}
\end{align} 
The stereographic projections can be easily determined from its definition and we find that there are several parametrizations of these projections. For example:
\begin{align}
\pi_2 (x_1, x_2, x_3 ) & = 
\left(
\frac{x_1}{1 - x_3} \, ,
\frac{x_2}{1 - x_3}
\right)
\, ,
\label{stereo-S2}
\\
\pi_3 (x_0 , x_1, x_2, x_3 ) & = 
\left(
\frac{x_1}{1 - x_0} \, ,
\frac{x_2}{1 - x_0} \, ,
\frac{x_3}{1 - x_0} 
\right) \, .
\label{stereo-S3}
\end{align}
The Hopf map inverse $\gamma = h^{-1}$ takes points $(x,z)$ from $S^2$ into loops on $S^3$. In general, if the points are different $(x,z) \neq (x', z')$, so are the corresponding loops $\gamma \neq \gamma'$. However, there is an object associated to a pair of loops called the \emph{Hopf invariant} of $h$ that is the linking number of the pair of loops, also called the \emph{Hopf index} of $(\gamma, \gamma')$, and which has the following general form
\begin{equation}
H(h) = l(\gamma , \gamma' ) \, .
\label{Hopf-index}
\end{equation}
$H(h)$ is an homotopy invariant that characterizes the Hopf map.
It is useful to write the Hopf invariant in terms of the Chern-Simons integral \cite{Whitehead:1947} of some vector fields. This can be done as follows. Consider an unit vector field $\mathbf{U}(\mathbf{x})$ with the following properties
\begin{equation}
\delta_{ij} U_i(\mathbf{x})  U_j(\mathbf{x})  = 1 \, ,
\qquad |\mathbf{U} (\mathbf{x})| \to \mathbf{u} \, 
\, \mbox{if} \, \, |\mathbf{x}| \to \infty \, ,
\label{unit-vector-field}
\end{equation}
where $\mathbf{u}$ is a constant unit vector. Then define the vector field  $\mathbf{F}(\mathbf{x})$ with the following components
\begin{equation}
F_j (\mathbf{x}) = \varepsilon_{jmn} \varepsilon_{prs}
U_p (\mathbf{x}) \partial_m U_r (\mathbf{x}) 
\partial_n  U_s (\mathbf{x}) \, .
\label{F-field-from-U}
\end{equation}
The field  $\mathbf{F}(\mathbf{x})$ can be written in terms of a potential vector field 
$\mathbf{A}(\mathbf{x})$ as follows
\begin{equation}
F_j (\mathbf{x}) = \varepsilon_{jmn}\partial_m A_n (\mathbf{x}) \, .
\label{F-field-potential}
\end{equation}
Then the Hopf index of the field $\mathbf{U}(\mathbf{x})$ is given by the following equalities
\begin{equation}
H(\mathbf{U}) = \int d^3 x \delta_{ij} F_i (\mathbf{x})  A_j (\mathbf{x}) 
= \int d^3 x  \varepsilon_{jmn} A_j (\mathbf{x}) \partial_m A_n (\mathbf{x}) 
\, .
\label{Hopf-index-integral}
\end{equation}
We recognize in the last equality above the explicit form of the Chern-Simons integral for the potential $\mathbf{A}(\mathbf{x})$.


\newpage


\begin{thebibliography}{Proper}

\bibitem{Trautman:1977im} 
  Trautman, A. (1977). ``Solutions of the Maxwell and Yang-Mills Equations Associated with Hopf Fibrings,''
  \emph{Int. J. Theor. Phys.} 16, 561.

\bibitem{Ranada:1989wc} 
  Ra\~{n}ada, A. F. (1989).
  ``A Topological Theory of the Electromagnetic Field,''
  \emph{Lett. Math. Phys.} 18, 97.

\bibitem{Ranada:1990}
  Ra\~{n}ada,  A. F. (1990).
``Knotted solutions of the Maxwell equations in vacuum,''
  \emph{J. Phys. A} 23 L815. 

\bibitem{Ranada:1992hw} 
  Ra\~{n}ada,  A. F. (1992).
  ``Topological electromagnetism,''
  \emph{J. Phys. A} 25, 1621.

\bibitem{Ranada:1995}
  Ra\~{n}ada, A. F. and Trueba, J. L. (1995).
  ``Electromagnetic Knots,''
  \emph{Phys. Lett. A} 202, 337-342.

\bibitem{Ranada:1997}
  Ra\~{n}ada,  A. F. and Trueba, J. L. (1997).
  ``Two properties of electromagnetic knots,''
  \emph{Phys. Lett. A} 222, 25-33.

 

\bibitem{Irvine:2008}
  Irvine, W. T. M. and Bouwmeester, D. (2008).
  ``Linked and knotted beams of light,''
  \emph{Nature Physics} 4, 716 - 720.

\bibitem{Irvine:2010}
  Irvine, W. T. M. (2010).
  ``Linked and knotted beams of light, conservation of helicity and the flow of null electromagnetic fields,''
  \emph{J. Phys. A: Math. Theor.} 43 385203.


\bibitem{Arrayas:2010xi} 
  Array\'{a}s, M. and Trueba, J. L. (2010).
  ``Motion of charged particles in a knotted electromagnetic field,''
  \emph{J. Phys. A} 43, 235401.


\bibitem{Kleckner:2013}
  Kleckner, D. and Irvine, W. T. M. (2013).
  ``Creation and dynamics of knotted vortices,''
  \emph{Nature Physics} 9, 253-258.

\bibitem{Arrayas:2011ci} 
  Array\'{a}s, M. and Trueba, J. L. (2012).
  ``Exchange of helicity in a knotted electromagnetic field,''
  \emph{Annalen Phys.}  524, 71.

\bibitem{Arrayas:2017wtr} 
  Array\'{a}s, M. and Trueba, J. L. (2017).
  ``Collision of two hopfions,''
  \emph{J. Phys. A} 50, 085203.
	
\bibitem{Ranada:2017ore} 
Ra\~{n}ada,  A. F., Tiemblo, A. and Trueba, J. L. (2017). 
  ``Time evolving potentials for electromagnetic knots,''
  \emph{Int. J. Geom. Meth. Mod. Phys.} 14, 1750073.


\bibitem{Ranada:1998vp} 
  Ra\~{n}ada,  A. F. and Trueba, J. L. (1998).
  ``A Topological mechanism of discretization for the electric charge,''
  \emph{Phys. Lett. B} 422, 196.

\bibitem{Ranada:2003}
Ra\~{n}ada,  A. F. (2003). 
  ``Interplay of topology and quantization: topological energy quantization in a cavity,''
  \emph{Phys. Lett. A} 310, 434-444.

\bibitem{Ranada:2006tq} 
  Ra\~{n}ada,  A. F. and Trueba, J. L. (2006).
  ``Topological quantization of the magnetic flux,''
  \emph{Found. Phys.} 36, 427.

\bibitem{Arrayas:2012eja} 
  Array\'{a}s, M., Trueba, J. L. and Ra\~{n}ada,  A. F. (2012). 
  ``Topological Electromagnetism: Knots and Quantization Rules,''
  in \emph{Trends in Electromagnetism - From Fundamentals to Applications},
  Barsan, V. and Lungu, R. P. (Eds.), IntechOpen Ltd.


\bibitem{Arrayas:2011ia} 
  Array\'{a}s, M. and Trueba, J. L. (2011).
  ``Electromagnetic Torus Knots,''
  \emph{J. Phys. A: Math. Theor.} 48, 025203.

\bibitem{Kedia:2013bw} 
  Kedia, H., Bialynicki-Birula, I., Peralta-Salas, D. and Irvine,
  W. T. M. (2013).
  ``Tying knots in light fields,''
  \emph{Phys. Rev. Lett.}  111, 150404.

\bibitem{Hoyos:2015bxa}
  Hoyos, C., Sircar, N. and Sonnenschein, J. (2015). 
  ``New knotted solutions of Maxwell's equations,''
  \emph{J. Phys. A} 48, no. 25, 255204.


\bibitem{Goulart:2016orx} 
  Goulart, \'{E}. (2016). 
  ``Nonlinear electrodynamics is skilled with knots,''
  \emph{Europhys. Lett.}  115, 10004.

\bibitem{Alves:2017ggb} 
  Alves, D. W.F., Hoyos, C., Nastase, H. and Sonnenschein, J.
  (2017). 
  ``Knotted solutions for linear and nonlinear theories: electromagnetism and fluid dynamics,''
  \emph{Phys. Lett. B} 773, 412.
  
\bibitem{Alves:2017zjt} 
  Alves, D. W. F., Hoyos, C., Nastase, H. and Sonnenschein, J. (2017).
  ``Knotted solutions, from electromagnetism to fluid dynamics,''
  \emph{Int. J. Mod. Phys. A} 32, 1750200.


\bibitem{Ranada:1996}
  Ra\~{n}ada,  A. F. and Trueba, J. L. (1996).
  ``Ball lightning an electromagnetic knot,''
  \emph{Nature} 383, 32.

\bibitem{Irvine:2014}
  Irvine, W. T. M. and Kleckner, D. (2014).
  ``Liquid crystals: Tangled loops and knots,''
  \emph{Nature Materials} 13, 229–231. 

\bibitem{Smiet:2015} 
Smiet, C. B., Candelaresi, S., Thompson, A., Swearngin, J., Dalhuisen, J. W. and Bouwmeester, D. (2015). 
``Self-Organizing Knotted Magnetic Structures in Plasma,''
  \emph{Phys. Rev. Lett.} 115, 095001.



\bibitem{Ren:2008zzf}
  Ji-Rong, R., Tao, Z. and Shu-Fan, M. (2008). 
  J.~R.~Ren, T.~Zhu and S.~F.~Mo,
  ``Knotted topological phase singularities of electromagnetic field,''
  \emph{Commun. Theor. Phys.}  50, 1071.

\bibitem{deKlerk:2017qvq} 
 de Klerk, A. J. J. M., van der Veen, R. I., Dalhuisen, J. W. and Bouwmeester, D. (2017).  
  ``Knotted optical vortices in exact solutions to Maxwell's equations,''
  \emph{Phys. Rev. A} 95, 053820.
  

\bibitem{Trueba:2008sc}
Trueba, J. L. (2008).
``Electromagnetic knots and the magnetic flux in superconductors,''
\emph{Ann. Fond. Louis de Broglie} 33, 183-192.


\bibitem{Arrayas:2017sfq} 
  Array\'{a}s, M., Bouwmeester, D. and Trueba, J. L. (2017).
  ``Knots in electromagnetism,''
  \emph{Phys. Rept.}  667.




\bibitem{Kopinski:2017nvp} 
  Kopi\'{n}ski, J. and Nat\'{a}rio, J. (2017).
  ``On a remarkable electromagnetic field in the Einstein Universe,''
  \emph{Gen. Rel. Grav.} 49, 81.


\bibitem{Vancea:2017tmx} 
  Vancea, I. V. (2017).
  ``On the existence of the field line solutions of the Einstein -
  Maxwell equations,''
  \emph{Int. J. Geom. Meth. Mod. Phys.}  15, 1850054.


\bibitem{Silva:2018ule} 
  Costa e Silva, W., Goulart, E. and Ottoni, J. E. (2018).
  ``On spacetime foliations and electromagnetic knots,''
  arXiv:1809.09259 [math-ph].


\bibitem{Alves:2018wku} 
Alves, D. W. F. and Nastase, H. (2018).
 ``Hopfion solutions in gravity and a null fluid/gravity conjecture,''
  arXiv:1812.08630 [hep-th].


\bibitem{Lechtenfeld:2017tif} 
  O.~Lechtenfeld and G.~Zhilin,
  ``A new construction of rational electromagnetic knots,''
  Phys.\ Lett.\ A \textbf{382}, 1528 (2018).
  \doi{10.1016/j.physleta.2018.04.027}

\bibitem{Kumar:2020xjr} 
  O.~Lechtenfeld and K.~Kumar,
  ``On rational electromagnetic fields,''
  arXiv:2002.01005 [hep-th].



\bibitem{Vancea:2019yjt} 
  I.~V.~Vancea,
  ``Field Line Solutions of the Einstein-Maxwell Equations,''
   in \emph{An Essential Guide to Maxwell's Equations}, Ed. Casey
   Erickson, Nova Science Publishers.


\bibitem{Jackson:1998nia} 
  Jackson, J. D. (1962).
  ``Classical Electrodynamics,''
  John Wiley and Sons.

\bibitem{Frankel:1997ec} 
  Frankel, T. (1997).
  ``The geometry of physics: An introduction,''
  Cambridge University Press.

\bibitem{Milnor:1997} 
Milnor, J. (1997).
``Topology from the Differentiable Viewpoint,''
Princeton University Press.

\bibitem{Bateman:2016} 
Bateman, H. (2016).
``The Mathematical Analysis of Electrical and Optical Wave-Motion,''
Cambridge University Press.


\bibitem{Thompson:2014owa}
Thompson, A., Wickes, A., Swearngin, J. and Bouwmeester, D. (2015). 
  ``Classification of Electromagnetic and Gravitational Hopfions by Algebraic Type,''
  \emph{J. Phys. A} 48, 205202.

\bibitem{Kedia:2016nwk} Kedia, H., Foster, D., Dennis, M. R. and Irvine, W. T. M. (2016). ``Weaving knotted vector fields with tunable helicity,''
  \emph{Phys. Rev. Lett.} 117, 274501.

\bibitem{Dennis:2010}
 Dennis, M. R., King, R. P., Jack, B., O'Holleran, K. and Padgett , M. J. (2010).
   ``Isolated optical vortex knots,''
   \emph{Nature Physics} 6, 118-121.

\bibitem{Whitehead:1947} 
  Whitehead, J. H. C. 1947.
  ``An expression of Hop's invariant as an integral,''
  \emph{Proc. Nat. Acad. Sci. U.S.A.} 33, 117–123.

\end{thebibliography}
\end{document}